\documentclass[floatfix]{revtex4}

\usepackage{epsfig}
\usepackage{graphicx}
\begin{document}
\bibliographystyle{revtex}


\title{Update of Discovery Limits for Extra Neutral Gauge Bosons at 
Hadron Colliders}

\author{Stephen Godfrey}
\email[]{godfrey@physics.carleton.ca}
\affiliation{Ottawa-Carleton Institute for Physics \\
Department of Physics, Carleton University, Ottawa, Canada K1S 5B6}

\date{\today}

\begin{abstract}
We study and compare the discovery potential 
for heavy neutral gauge bosons ($Z'$) at the various hadron 
colliders under discussion at Snowmass 2001 which range in $\sqrt{s}$ from 
14~TeV for the LHC to 200~TeV for a variant of the VLHC.  
Typical search limits for $pp$ colliders 
are  $\sim 0.25-0.30 \times \sqrt{s}$ assuming $100$~fb$^{-1}$ to
$1$~ab$^{-1}$ of integrated luminosity with some 
variation due to differences of fermion couplings in the different 
models.  Discovery limits at the Tevatron are $\sim 1$~TeV for 
$15$~fb$^{-1}$, approximately 30--50\% higher than
this rough guideline, due to the higher 
$q\bar{q}$ luminosities in the $p\bar{p}$ beams.
\end{abstract}

\maketitle


Extended gauge symmetries and the associated heavy neutral gauge 
bosons, $Z'$, are a feature of many extensions of the standard model 
such as grand unified theories, Left-Right symmetric models, and 
superstring theories.  If a $Z'$ were discovered it would have 
important implications for what lies beyond the standard model.  
It is therefore important to study and compare the discovery 
reach for extra gauge bosons at the various facilities 
that are under consideration for the future
\cite{cvetic,godfrey,capstick,leike,riemann,rizzo}. 
Included in the list of 
proposed facilities considered at the Snowmass'01
workshop are high energy hadron colliders. 
In this report we update previous studies 
\cite{cvetic,godfrey,capstick,rizzo}
to include the high energy hadron colliders discussed at this meeting
which range in $\sqrt{s}$ from 14~TeV to 200~TeV.


Many models that predict extra gauge bosons exist in the 
literature.  We present search limits for several of these models 
that have received recent attention. Although far 
from exhaustive, the list forms
a  representative set for the purposes of comparison.  
The Effective Rank 5 E6 Model 
starts with the GUT group $E_6$ and breaks via the chain 
$E_6 \to SO(10)\times U(1)_\psi \to SU(5)\times U(1)_\chi \times 
U(1)_\psi$ with the $Z'$ charges given by linear combinations of the 
$U(1)_\chi$ and $U(1)_\psi$ charges.
Specific models are $Z_\chi$ 
corresponding to the extra $Z'$ of $SO(10)$, $Z_\psi$ 
corresponding to the extra $Z'$ of $E_6$, 
and $Z_\eta$ 
corresponding to the extra $Z'$ arising in some superstring theories.
The Left-Right symmetric model (LRM) is based on the gauge group 
$SU(2)_L \times SU(2)_R \times U(1)_{B-L}$,
which has right-handed charged currents and restores parity at high 
energy.
The Alternative Left-Right Symmetric model (ALRM) is based on the same 
gauge group as the LRM but now arising from $E_6$ where the fermion 
assignments are different  from those of the LRM 
due to an ambiguity in how they are embedded in the {\bf 27} 
representation..
The Un-Unified model (UUM) 
is based on the gauge group $SU(2)_q \times SU(2)_l 
\times U(1)_Y$ where the quarks and leptons each transform under their own 
$SU(2)$
and the KK model (KK) 
contains the Kaluza-Klein excitations of the SM gauge bosons that
are a possible consequence of theories with large extra dimensions. We 
also consider a $Z'$ with SM couplings (SSM). Details of these models 
and references are given in Ref. \cite{cvetic}.


The signal for a $Z'$ at a hadron collider consists of 
Drell-Yan production of lepton pairs 
\cite{capstick,godfrey,guts,pp,ehlq} 
with high invariant mass via 
$p p \to Z' \to l^+ l^-$.
The cross section for the production of on-shell $Z'$s is given by 
\cite{capstick}:
\begin{equation}
{\hbox{d} \sigma (pp\rightarrow f\bar f) \over \hbox{d} y}
={x_A x_B \pi^2 \alpha_{em}^2 (g_{Z'}/g_{Z^0})^4 \over 9 M_{Z'} 
\Gamma_{Z'} }\left({C_L^f}^2 +{C_R^f}^2 \right) \sum_q
\left({C_L^q}^2 +{C_R^q}^2 \right) G_q^+(x_A,x_B,Q^2)
\end{equation}
where
\begin{equation}
G_q^+(x_A,x_B,Q^2)=\sum_q \left[f_{q/A}(x_A) f_{\bar q/B}(x_B) +
f_{\bar q/A}(x_A) f_{q/B}(x_B) \right]
\end{equation}

The cross section for $Z'$ production at hadron colliders
is inversely proportional to the $Z'$  width.  If exotic
decay modes are kinematically allowed, the $Z'$ width will become larger and
more significantly, the branching ratios to conventional fermions smaller.  
We will only consider the case that no new decay modes are allowed.
The partial widths are given (at tree level) by 
\begin{equation}
\Gamma_{Z' \rightarrow f \overline f } =
M_{Z'} g^2_{Z'} (C'^2_{f_L} +C'^2_{f_R})\big/24\pi
\end{equation}

\begin{figure}
\includegraphics[width=11cm]{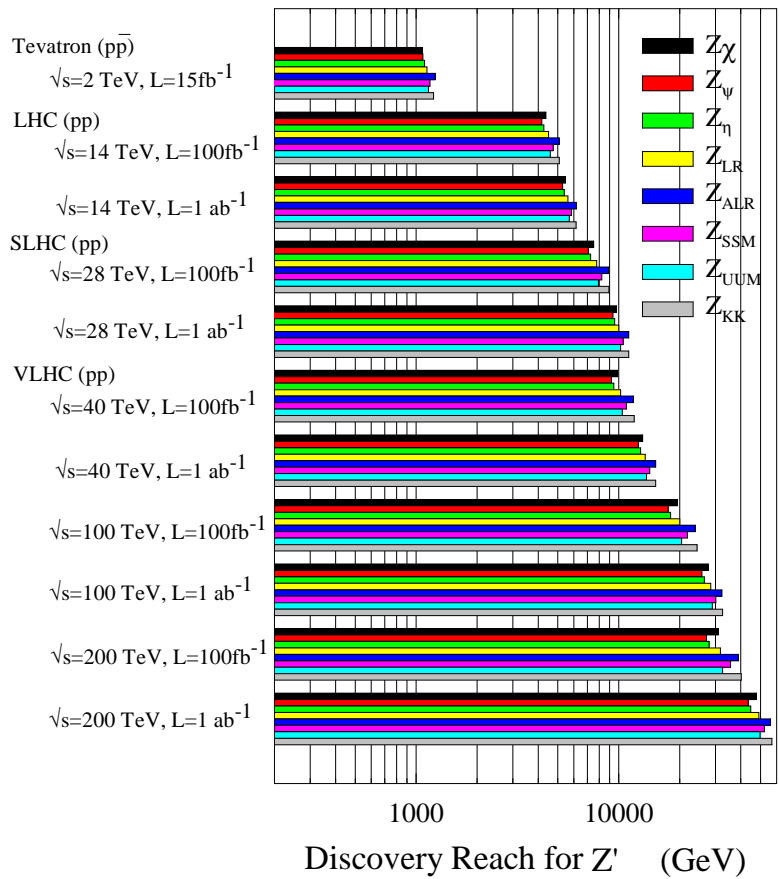} 
\caption{
Discovery limits for extra neutral gauge bosons ($Z'$) 
for the models described in the text based  
on 10 events in the $e^+e^-\ +\ \mu^+\mu^-$ channels.
}
\label{Fig2}
\end{figure}

We obtain the discovery limits for this process based on 10 events 
in the $e^+e^- + \mu^+\mu^-$ channels using the EHLQ quark 
distribution functions \cite{ehlq} set 1, taking $\alpha=1/128.5$, 
$\sin^2\theta_w=0.23$, and including a 1-loop $K$-factor in the $Z'$ 
production \cite{k-factor}. 
We include 2-loop QCD radiative corrections and 1-loop QED radiative 
corrections in calculating the $Z'$ width.  Using different 
quark distribution functions results in a roughly 10\% variation in 
the $Z'$ cross sections \cite{rizzo} with the subsequent change in 
discovery limits.  Detailed 
detector simulations for the Tevatron and LHC validated our 
approximations as a good estimator of the true search reach.  
Furthermore, the results of our previous studies following this 
approach are totally 
consistent with subsequent experimental limits obtained at the 
Tevatron.

Lowering the number of events in the $e^+e^- + \mu^+\mu^-$ channels 
to 6 raises the discovery reach  about $10\%$ while lowering the
luminosity by a factor of ten  reduces the reach  by about a factor 
of 3 \cite{capstick}.

In our calculations we assumed that the $Z'$ only decays into the 
three conventional fermion families.  If other decay channels were 
possible, such as to exotic fermions filling out larger fermion 
representations or supersymmetric partners, the $Z'$ width would be 
larger, lowering the discovery limits.  On the other hand, if decays 
to exotic fermions were kinematically allowed, the 
discovery of exotic fermions would be an important discovery in 
itself;  the study of the corresponding decay modes would provide additional
information on the  nature of the extended gauge structure.   

The discovery limits for various models at hadron colliders 
are shown in Fig. 1.
These  bounds are relatively insensitive to specific models.  In addition, 
since they are based on a distinct signal with little background 
they are relatively robust limits.  
Typical search limits for $pp$ colliders 
are  $\sim 0.25-0.30 \times \sqrt{s}$ assuming $100$~fb$^{-1}$ to
$1$~ab$^{-1}$ of integrated luminosity with some 
variation due to differences of fermion couplings in the different 
models.
The Tevatron, a $p\bar{p}$ collider, has a 50\% higher discovery
reach than this rough guideline, 
indicating that valence quark contributions
to the Drell-Yan production process are still important at these
energies.

\begin{acknowledgments}
This work was supported by the Natural Sciences and 
Engineering Research Council of Canada.  The author thanks JoAnne 
Hewett for her encouragement.
\end{acknowledgments}



\end{document}